\documentclass[prl,10pt,twocolumn,superscriptaddress,showpacs]{revtex4}
\usepackage{amsmath}
\usepackage{latexsym}
\usepackage{amssymb}
\usepackage{graphics,epstopdf}
\usepackage{graphicx}
\usepackage[colorlinks=true, citecolor=blue, urlcolor=blue ]{hyperref}
\usepackage{float}
\usepackage{graphicx}
\usepackage{amsfonts}

\begin{document}

\title{Steering a single system sequentially by multiple observers}

\author{Souradeep Sasmal}
\email{souradeep@mail.jcbose.ac.in}
\affiliation{Centre for Astroparticle Physics and Space Science (CAPSS),
Bose Institute, Block EN, Sector V, Salt Lake, Kolkata 700 091, India}

\author{Debarshi Das}
\email{debarshidas@jcbose.ac.in}
\affiliation{Centre for Astroparticle Physics and Space Science (CAPSS),
Bose Institute, Block EN, Sector V, Salt Lake, Kolkata 700 091, India}

\author{Shiladitya Mal}
\email{shiladitya.27@gmail.com}
\affiliation{S. N. Bose National Centre for Basic Sciences, Block JD, Sector III, Salt Lake, Kolkata 700 098, India}
\affiliation{Harish-Chandra Research Institute, HBNI, Chhatnag Road, Jhunsi, Allahabad 211 019, India}

\author{A. S. Majumdar}
\email{archan@bose.res.in}
\affiliation{S. N. Bose National Centre for Basic Sciences, Block JD, Sector III, Salt Lake, Kolkata 700 098, India}

\begin{abstract}

Quantum mechanics puts a restriction on the number of observers who can simultaneously steer another observer's system, known as the monogamy of steering. In this work we find the limit of the number of observers (Bobs) who can steer another party's (Alice's) system invoking a scenario where half of an entangled pair is shared between a single Alice in one wing and several Bobs on the other wing, who 
act sequentially and independently of each other. When all the observers measure two dichotomic observables, we find that two Bobs can steer Alice's system going beyond the monogamy restriction. We 
further show that three Bobs can steer Alice's system considering a three-settings linear steering inequality, and then conjecture that at most $n$ Bobs can demonstrate steering of Alice's system when steering is probed through an $n$-settings linear steering inequality.
\end{abstract} 


\maketitle

\section{I. INTRODUCTION} 

The subject of quantum correlations first received attention due to the seminal paper by Einstein, Podolsky and Rosen (EPR) \cite{EPR}, leading subsequently to the exploration by Bell that a local realist description which reflects classical mechanics is incompatible with quantum mechanics (QM) \cite{bell}. Schr\"odinger unlike  EPR, did not believe in the incompleteness of QM, but he was surprised by the fact that an observer  can steer a system which is not in her possession, and conjectured that such an odd event cannot be experimentally verified \cite{sr}. Many years later, Reid proposed a criterion for experimentally demonstrating the   EPR argument \cite{reid} using 
the Heisenberg uncertainty relation.

 Later, Wiseman et. al. \cite{wiseman, wiseman2} have  shown  that steering is the lack of a local hidden state (LHS) model through which an observer can simulate classically another remote party's state. It has been shown that the Reid criterion can be derived from the existence of a LHS model \cite{cvl}. 
  Quantum correlations have many information theoretic applications, such as sharing of secret key by distant parties \cite{bb84, ekert}, teleportation \cite{tele}, dense coding \cite{dens}, 
 and remote state preparation \cite{pati}. Exploration of new properties of quantum correlations is of utmost importance not only for information theoretic purposes, but also for quantum foundational aspects.

A key property of all quantum correlations is that they are monogamous in nature. The restriction on sharing of quantum correlations  between several number of observers is quantitatively expressed through monogamy relations for entanglement \cite{ckw}, Bell-nonlocality  \cite{tv}, and EPR steering \cite{mreid, achshsm, mono1_n, mono2_n}. Unlike entanglement and Bell-nonlocality, EPR steering is asymmetric with respect to the observers resulting in extra sophistication in the monogamy relations. Reid derived monogamy constraints contingent upon using some steering witnesses based on inferred variances and showed that they have a directional property \cite{mreid}. In particular, two parties cannot independently steer a third party's system using the same two-settings steering witnesses. However, the converse is not true, i. e., it is possible for one party to steer the other two systems. Recently, a monogamy constraint has been derived using the volume of the quantum steering ellipsoid, which states that one party cannot steer both of the other two parties to a large set of states \cite{vm}. In \cite{mreid} it was also shown that for multipartite systems, at most $n-1$ parties can steer the $n$th party's system using an $n$-observable steering inequality. 

Quantum correlations also satisfy a fundamental physical postulate, i. e., the no-signalling condition (the probability of obtaining one party's outcome does not depend on the other spatially separated party's setting). Recently, it has been  shown in \cite{sygp} that monogamy relations no longer hold for the correlations obtained from measurements performed by different parties if the no-signalling condition is relaxed.  Here,  relaxing the no-signalling condition does not imply violating relativistic causality, but it implies a scenario where sequential measurements are performed on the same particle by different observers. In this scenario all the marginal probabilities of subsets of observers can not be determined from the joint probability of outcomes of the measurements performed by all observers. We explain this issue in more details later. Specifically, the scenario is that one observer (say, Alice) has access to half of an entangled pair whereas several observers (say, several Bobs) can access and measure on another half of that pair sequentially. It has been shown through numerical evidences that at most two Bobs can demonstrate violation of the Clauser-Horne-Shimony-Holt (CHSH)  inequality \cite{chsh} with single Alice when the measurements of each of the several Bobs at one side are unbiased with respect to the previous Bobs. This result is proved analytically in \cite{mal} and confirmed by  experiment \cite{ex1, ex2}.

In the present work we investigate how many Bobs can steer single Alice, when Alice shares half of a bipartite entangled state, and multiple Bobs share and measure sequentially on another half. In order to address this issue we consider the analog of the CHSH inequality for steering, which is a necessary and sufficient condition for steering in the experimental scenario where each party measures two dichotomic observables with mutually unbiased measurements for the party whose system is being steered \cite{achsh}. In this case we find that at most two Bobs can steer Alice's system.
 We also consider linear steering inequalities with two as well as three settings which was proposed in \cite{cvl}. Maximum quantum violation of this linear inequality was derived in \cite{costa}, and a monogamy relation for this inequality was also derived in \cite{mreid}. In the scenario of a single Alice and multiple Bobs \cite{sygp, mal}, we find that at most two Bobs can steer  Alice via violation of the two settings linear steering inequality. On the other hand, at most three Bobs can steer  Alice via violation of the three settings linear steering inequality. Based on the above result, we conjecture that when steering is probed through $n$-settings linear steering inequality, at most $n$ Bobs can steer Alice's system.
 
We organize this paper in the following way. In Sec. II, we discuss in brief EPR steering in the
context of two separate steering inequalities which we apply later. Next, in Sec. III, we consider the scenario where Alice shares half of a bipartite entangled state, and multiple Bobs share and measure sequentially on another half. In this context we obtain a bound on the number of Bobs who can steer
a single Alice determined through the quantum violations of the steering inequalities considered. We end with concluding remarks in Sec. IV.

\section{II. Different steering inequalities}

 Wiseman et al. \cite{wiseman} have shown  that Alice can steer Bob's system if there is no fixed ensemble known to Alice by which she can simulate different states for Bob pertaining to different measurement choices. In other words, there would be no local hidden variable-local hidden state (LHV-LHS) model for the correlation obtained from the local measurements on the shared bipartite state.
 
 Suppose $A \in \mathbb{F}_{\alpha} $ and $B \in \mathbb{F}_{\beta}$ are the possible choices of measurements for two spatially separated observers, say Alice and Bob, with outcomes $a \in \mathbb{G}_{a}$ and $b \in \mathbb{G}_{b}$, respectively. The joint probability of obtaining the outcomes $a$ and $b$, when measurements $A$ and $B$ are performed by Alice and Bob locally on state $\rho_{AB}$, respectively, is given by, $P(a,b|A,B,\rho_{AB})$.
The bipartite state $\rho_{AB}$ of the system is steerable by Alice to Bob iff it is not the case that for all  $A \in \mathbb{F}_{\alpha} $, $B \in \mathbb{F}_{\beta}$, $a \in \mathbb{G}_{a}$, $b \in \mathbb{G}_{b}$, the joint probability distribution can be written in the form
\begin{equation}
P(a, b|A, B, \rho_{AB}) = \sum_{\lambda} P(\lambda )P(a|A,\lambda)P_Q(b|B,\rho_{\lambda})
\end{equation}
where $P(\lambda)$ is the probability distribution over the hidden variables $\lambda$, and $P(a|A,\lambda)$ denotes an arbitrary probability distribution and $P_Q(b|B,\rho_{\lambda})(=tr [\rho_{\lambda}\mathbb{Q}^b_B])$ denotes the quantum probability of outcome $b$ given measurement $B$ on the local hidden state $\rho_{\lambda}$; $\mathbb{Q}^b_B$ being the measurement operator of the observable $B$ associated with outcome $b$.

The necessary and sufficient criterion \cite{achsh} to detect steering from Bob to Alice in a scenario where both  observers have the choice to perform between two dichotomic measurements ($\{ A_1, A_2 \}$ for Alice and $\{ B_1, B_2 \}$ for Bob) with Alice's measurements being mutually unbiased, is given by,
\begin{eqnarray}\label{si}
S_{BA}=\sqrt{\langle (B_1+B_2) A_1\rangle^2+\langle (B_1+B_2) A_2\rangle^2}+\nonumber\\
\sqrt{\langle (B_1- B_2) A_1\rangle^2+\langle (B_1- B_2) A_2 \rangle^2}\leq 2.
\end{eqnarray}
This inequality is the analog of the CHSH inequality for steering. We will henceforth call it 
the Cavalcanti-Foster-Fuwa-Wiseman (CFFW)  \cite{achsh} inequality.

In \cite{cvl} the authors have developed a series of steering inequalities when both the parties are allowed to perform $n$ number of dichotomic measurements
on his or her part of the subsystem. The inequalities have the following form:
\begin{equation}
F^n = \dfrac{1}{\sqrt{n}} \Big| \sum_{i=1}^{n} \langle A_i \otimes B_i \rangle \Big| \leq 1,
\end{equation}
where, $A_i= \hat{u_i}\cdot\vec{\sigma}$, $B_i= \hat{v_i}\cdot\vec{\sigma}$, $\vec{\sigma}=(\sigma_1, \sigma_2, \sigma_3)$ is a vector composed of Pauli matrices, $\hat{u_i} \in \mathbb{R}^3$ are unit vectors. For $n$ equals to 2 or 3, $\hat{v_i} \in \mathbb{R}^3$ are orthonormal vectors. $\langle A_i \otimes B_i \rangle$ = Tr$(\rho_{AB}  A_i \otimes B_i)$ with $\rho_{AB} \in \mathcal{H}_A \otimes \mathcal{H}_B$ being the bipartite quantum system shared between the two parties. We will call this inequality as the
 Cavalcanti-Jones-Wiseman-Reid (CJWR) \cite{cvl} inequality henceforth.

\section{III. PROBING QUANTUM STEERING WITH MULTIPLE OBSERVERS AT ONE SIDE}

A monogamy relation for the CFFW inequality derived in \cite{achshsm} implies that for three parties sharing a quantum state, any two of them cannot steer the third party's state simultaneously. In the context of the CJWR inequality, it is  known \cite{mreid} that for spatially separated multiple observers sharing a quantum state, at most $n-1$ parties can steer the $n$th party's system using $n$-settings CJWR inequality \cite{mreid}. In other words, more than $n-1$ observers cannot steer the $n$th party's state, which is probed through violation of the $n$-settings CJWR inequality.
We now show that there exists a scenario where two parties can steer a single party simultaneously, with two dichotomic measurements performed by each of the parties. In case of three dichotomic measurements performed by each of the parties, we show that three parties can steer a single party simultaneously.

 Monogamy relations derived in \cite{mreid, achshsm} are applicable in the scenario where 
 the no-signalling condition is satisfied between each pair of observers. Now consider a scenario where the no-signalling condition is relaxed for a subset of observers. Let us briefly describe the scenario as introduced in \cite{sygp}.
Suppose, two spin-$\frac{1}{2}$ particles are prepared in the singlet state given by,
\begin{equation}
|\psi \rangle = \frac{1}{\sqrt{2}} (|01 \rangle - |10 \rangle)
\end{equation}
where $|0\rangle$ and $|1\rangle$ are the eigenstates of $\sigma_z$ operator. These two particles are spatially separated. Alice measures on the first particle and multiple Bobs (Bob$^1$,  Bob$^2$, Bob$^3$, ..., Bob$^n$) measure on the second particle sequentially. After doing measurements on his particle Bob$^1$ delivers the particle to Bob$^2$, and similarly, Bob$^2$ passes his particle to Bob$^3$ after  measurement, and so on. Here, it is important to note that each Bob performs measurements independent of the measurement settings and outcomes of the previous Bobs on the particle in his possession. Moreover, we are considering 
an unbiased input scenario, i. e.,  all possible measurement settings of each Bob are equally probable. Here, the  no-signalling condition is not satisfied between each pair of observers (between different Bobs).

No-signalling condition, a consequence of relativistic causality,  is applicable between any set of spatially separated observers when each of the observers performs measurements on different  non-interacting particles. In case of the multipartite scenario considered in monogamy context, $n$ observers share $n$ number of particles, one particle per each observer, and all the observers are spatially separated. Hence, the no-signalling condition is satisfied between any pair of observers. On the other hand, in the present scenario considered by us, there are two particles, one is possessed by Alice and the other is possessed by multiple Bobs. Alice and multiple Bobs are spatially separated and, hence, the no-signalling condition is satisfied between Alice and any Bob. For example, suppose $P(a, b_1, b_2| x, y_1, y_2)$ denotes the probability of obtaining the outcomes $a$, $b_1$ and $b_2$ when Alice, Bob$^1$ and Bob$^2$ perform measurements $x$, $y_1$ and $y_2$ respectively. Since, as a consequence of relativistic causality, no-signalling is satisfied between Alice and any Bob, we can write,
\begin{equation}
\sum_a P(a, b_1, b_2| x, y_1, y_2) = P(b_1, b_2|y_1, y_2).
\end{equation} 
However, in this case, different Bobs act sequentially on the same particle. In fact, Bob$^1$ implicitly signals to Bob$^2$ by his choice of measurement on the state before he passes it on, and similarly, Bob$^2$ signals to Bob$^3$ and so on. In other words, one can write the following:
\begin{equation}
\sum_{b_1} P(a, b_1, b_2| x, y_1, y_2) \neq P(a, b_2|x, y_2).
\end{equation}
The above equation indicates relaxing of the no-signalling condition defined in a broader framework pertinent to the scenario considered here, although relativistic causality is not violated here.
Note that Bob$^2$ performs measurement after Bob$^1$ performs his measurement on the particle. Hence, outcomes of the measurement by Bob$^1$ do not depend on the choice of measurement by Bob$^2$. Moreover, outcomes of Alice's measurement do not depend on the choice of measurement by Bob$^2$ due to the no-signalling condition. Mathematically, one can write the following:
\begin{equation}
\sum_{b_2} P(a, b_1, b_2| x, y_1, y_2) = P(a, b_1|x, y_1).
\end{equation}

 Against the above backdrop, we investigate how many Bobs can steer Alice's particle.
As we want to explore how many Bobs can have measurement statistics violating the CFFW and CJWR inequalities with a single Alice, each of the Bobs except the last Bob cannot measure sharply. If any Bob
measures sharply, there would be no possibility of violation of the CFFW or CJWR inequality by the next Bob,
since  the entanglement of the state shared between Alice and 
the subsequent Bob would be completely destroyed. Hence, in order to address the aforementioned problem with $n$ Bobs, the measurements of the first $(n -1)$ Bobs should be weak. 

\subsection{A. Weak measurement scenario}

For completeness of the paper here we briefly recapitulate the weak measurement scheme followed in \cite {sygp} and then the unsharp version of that considered in \cite{mal}.
In \cite {sygp} Bob's weak measurement is performed by a broad pointer state, and the optimal pointer state was employed to demonstrate the sharing of nonlocality in terms of the QM violation of the CHSH inequality  \cite{sygp}.   
In the standard Von Neuman measurement framework, after an interaction with a pointer having the state $\phi (q)$,  the state $| \psi \rangle (= a | 0 \rangle + b | 1 \rangle, |a|^2+|b|^2 = 1)$  of a spin-$\frac{1}{2}$ particle becomes
\begin{equation}
a | 0 \rangle \otimes \phi (q-1)  + b | 1 \rangle  \otimes \phi (q+1).
\end{equation}
The weak version of this ideal measurement is characterised by two parameters, namely the quality factor $F$ and the precision $G$ of the measurements. The quality factor is defined as $F(\phi) = \int_{-\infty}^{\infty} \langle \phi (q+1) | \phi (q-1) \rangle dq$. It quantifies the extent to which the state of the system remains undisturbed after the measurement. The precision of measurement is given by, $ G = \int_{-1}^{1} \phi ^2 (q) dq$. It quantifies the information gain from measurement. In case of strong measurement, $F = 0$ and $G = 1$.
An optimal pointer is defined as the one which gives the best trade-off between these two quantities, i.e., for a given quality factor, it provides the greatest precision. It has been shown that the information-disturbance trade-off condition for an optimal pointer is given by, $F^2 + G^2 =1$ \cite{sygp}.

This weak measurement formalism has been recast in the unsharp measurement formalism in Ref. \cite{mal}. Unsharp measurement is a particular class of positive operator valued measurement (POVM) \cite{pb1, pb2}. POVM is a set of positive operators that add up to identity, i.e.,  $E \equiv \{ E_i | \sum_i E_i = \mathbb{I}, 0 <E_i \leq \mathbb{I} \}$. 
The probability of getting the $i$-th outcome is Tr$[\rho E_i]$. Effects ($E_i$s) represent quantum events that may occur as outcomes of a measurement. In case of a dichotomic unsharp measurement, the effect operators are given by, 
\begin{equation}
E^\lambda_{\pm} = \lambda P_{\pm} + (1-\lambda) \frac{\mathbb{I}_2}{2},
\end{equation}
where $\lambda$ ($0 < \lambda \leq 1$) is the sharpness parameter, $P_{+}$ ($P_{-}$) is the projector associated with the outcome $+$ ($-$), $\mathbb{I}_2$ is the $2 \times 2$ identity matrix.
It was shown in \cite{mal} that weak measurement described by $F$ and $G$ is related to the unsharp measurement formalism through $F = \sqrt{1-\lambda^2}$ and $G = \lambda$. Hence, $\lambda$ characterizes the precision of the measurement. For $G = \lambda = 1$, $F$ becomes zero, this being the case of sharp measurement. Hence, in the unsharp measurement formalism, the optimal pointer state condition, $F^2 + G^2 = 1$, is naturally satisfied. 

\subsection{B. Two Bobs can steer Alice with two measurement settings for each party}

We assume that a maximally entangled state say, singlet state, is initially shared between Alice and multiple Bobs (say, Bob$^1$, Bob$^2$, Bob$^3$, ..., Bob$^n$). 
Suppose, Alice has a choice between two dichotomic measurements: spin component observables in the directions $\{ \hat{x^0}, \hat{x^1} \}$,  and Bob$^n$ also has the choice between the spin component observables in the directions $\{ \hat{y^0_n}, \hat{y^1_n} \}$. Outcomes of these measurements are labelled by $\{-1, +1 \}$.

The correlation function between Alice and Bob$^1$, when Alice performs a projective measurement of the spin component along the direction $\hat{x^j}$, and Bob$^1$ performs an unsharp measurement of the spin component along the direction $\hat{y_1^i}$, is given by,
\begin{equation}
C_1^{ji} = -\lambda_1 (\hat{y_1^i} \cdot \hat{x^j}).
\end{equation}
Now, for two Bobs measuring in succession, the joint probability of obtaining the outcomes $a$ and $b_2$ when Alice performs a projective measurement of the spin component along the direction $\hat{x^j}$ and Bob$^2$ performs an unsharp measurement of the spin component along the direction $\hat{y_2^k}$ respectively, given that Bob$^1$ has performed an unsharp measurement of the spin component along the direction $\hat{y_1^i}$, is given by,
\begin{equation}
p(a, b_2|x^j, y_1^i, y_2^k) = \frac{\sqrt{1 - \lambda_1^2}}{2} \frac{1 - a b_2 \lambda_2 (\hat{y_2^k} \cdot \hat{x^j})}{2} \nonumber
\end{equation}
\begin{equation}
+ \frac{1- \sqrt{1 - \lambda_1^2}}{2} \frac{1 - a b_2 \lambda_2 (\hat{y_1^i}\cdot \hat{x^j})(\hat{y_2^k}\cdot \hat{y_1^i})}{2},
\end{equation} 
where $\lambda_1$ and $\lambda_2$ are the sharpness parameters of the measurements done by Bob$^1$ and Bob$^2$, respectively.
The correlation function between Alice and Bob$^2$, in this case is given by,
\begin{align}
C_2^{jk} =& -\lambda_2 \Big[ \sqrt{1-\lambda_1^2} (\hat{y_2^k} \cdot \hat{x^j}) \nonumber\\
&+ \Big(1- \sqrt{1-\lambda_1^2} \Big)  (\hat{y_1^i} \cdot \hat{x^j})  (\hat{y_2^k} \cdot \hat{y_1^i})\Big].
\end{align}
Since Bob$^2$ is ignorant about the measurement settings of Bob$^1$, this correlation has to be averaged over the two possible measurement settings of Bob$^1$ (spin component observables in the directions $\{ \hat{y^0_1}, \hat{y^1_1} \}$). This average correlation function between Alice and Bob$^2$ is given by,
\begin{equation}
\overline{C_2^{jk}} = \sum_{i = 0,1} C_2^{jk} P(\hat{y_1^i}),
\end{equation}
where $P(\hat{y_1^i})$ is the probability with which Bob$^1$ performs measurement of spin component observables in the direction $\hat{y_1^i}$ ($i = 0, 1$). Since we are considering an unbiased input scenario, all the possible measurement settings of the previous Bob are equally probable, i. e.,  $P(\hat{y_1^0})$ = $P(\hat{y_1^1})$ = $\frac{1}{2}$. Hence, we obtain,
\begin{align}
\overline{C_2^{jk}} =& -\frac{\lambda_2}{2} \Big[\sqrt{1-\lambda_1^2} (\hat{y_2^k} \cdot \hat{x^j}) \nonumber \\
& + \Big(1- \sqrt{1-\lambda_1^2}\Big)  (\hat{y_1^0} \cdot \hat{x^j})  (\hat{y_2^k} \cdot \hat{y_1^0}) \Big] \nonumber \\
& -\frac{\lambda_2}{2} \Big[\sqrt{1-\lambda_1^2} (\hat{y_2^k} \cdot \hat{x^j}) \nonumber \\
&+ \Big(1- \sqrt{1-\lambda_1^2}\Big)  (\hat{y_1^1} \cdot \hat{x^j})  (\hat{y_2^k}\cdot \hat{y_1^1}) \Big]. 
\end{align}
In a similar way, the average correlation function between Alice and any Bob can be evaluated.

In this scenario, the necessary and sufficient condition for Bob$^n$ to steer Alice's particle, with Alice's measurement settings being mutually unbiased, is given by violation of the CFFW inequality \cite{achsh},
\begin{align}
S_n = & \sqrt{\Big(\overline{C_n^{00}} + \overline{C_n^{01}}\Big)^2 + \Big(\overline{C_n^{10}} + \overline{C_n^{11}}\Big)^2} \nonumber\\
& + \sqrt{\Big(\overline{C_n^{00}} - \overline{C_n^{01}}\Big)^2 + \Big(\overline{C_n^{10}} - \overline{C_n
^{11}}\Big)^2}  \le 2.
\label{chshst}
\end{align}
We assume that the two possible choices of measurement settings of Alice are the spin component observables in the directions $\hat{x^i}$ given by,
\begin{equation}
\label{alicedir}
\hat{x^i} = sin \theta^x_i cos \phi^x_i \hat{x} + sin \theta^x_i sin \phi^x_i \hat{y} + cos \theta^x_i \hat{z}.
\end{equation}
Similarly, the two possible choices of measurement settings of Bob$^n$ are the spin component observables in the directions $\hat{y_n^i}$ given by,
\begin{equation}
\label{bobndir}
\hat{y_n^i} = sin \theta^{y_n}_i cos \phi^{y_n}_i \hat{x} + sin \theta^{y_n}_i sin \phi^{y_n}_i \hat{y} + cos \theta^{y_n}_i \hat{z},
\end{equation}
where, $i\in\{0,1\}$.
As Alice's system is to be steered, her measurement settings should be mutually unbiased, i.e.,  $\hat{x^0} \cdot \hat{x^1} = 0$ \cite{achsh}.

Now consider whether Bob$^1$ and Bob$^2$ can steer Alice's system simultaneously. The measurements of the final Bob (i.e., Bob$^2$) are sharp ($\lambda_2 = 1$), and the measurements of Bob$^1$ are unsharp. We observe that when Bob$^1$ gets $5\%$ violation of the CFFW inequality for steering given by (\ref{chshst}), i.e., when $S_1 = 2.10$, then the maximum quantum mechanical violation of CFFW inequality for Bob$^2$ is $18\%$, i.e., $S_2 = 2.36$. This happens for the choice of measurement settings: ($\theta^x_0, \phi^x_0, \theta^x_1, \phi^x_1, \theta^{y_1}_0, \phi^{y_1}_0, \theta^{y_1}_1, \phi^{y_1}_1, \theta^{y_2}_0, \phi^{y_2}_0, \theta^{y_2}_1, \phi^{y_2}_1$) $\equiv$ $(\frac{\pi}{2}, 0, 0, 0, \frac{\pi}{4}, 0, \frac{3 \pi}{4}, 0,   \frac{\pi}{4}, 0, \frac{3 \pi}{4}, 0) $ with $\lambda_1 =0.74$. In fact, it is observed that for $\lambda_1 \in[0.71,0.91]$ both Bob$^1$ and Bob$^2$ can steer Alice. 

Next, we  address the question whether three Bobs, i.e., Bob$^1$, Bob$^2$ and Bob$^3$ can steer  Alice's particle. In this case, the measurements of the final Bob (i.e., Bob$^3$) are sharp ($\lambda_3 = 1$), and the measurements of Bob$^1$ and Bob$^2$ are unsharp. Here we observe that, when each of the QM violation of CFFW inequality for steering from Bob$^2$ to Alice and that for steering from Bob$^1$ to Alice is $5\%$, i.e., when $S_1 = 2.10$ and $S_2 = 2.10$, then the maximum QM value of the left hand side of the CFFW inequality for steering from Bob$^3$ to Alice is given by, $S_3 = 1.72$. This happens for the choice of measurement settings: ($\theta^x_0, \phi^x_0, \theta^x_1, \phi^x_1, \theta^{y_1}_0, \phi^{y_1}_0, \theta^{y_1}_1, \phi^{y_1}_1, \theta^{y_2}_0, \phi^{y_2}_0, \theta^{y_2}_1, \phi^{y_2}_1, \theta^{y_3}_0, \phi^{y_3}_0,$\\ 
$\theta^{y_3}_1, \phi^{y_3}_1$) $\equiv$ $(\frac{\pi}{2}, 0, 0, 0, \frac{\pi}{4}, 0, \frac{3 \pi}{4}, 0,   \frac{\pi}{4}, 0, \frac{3 \pi}{4}, 0,  \frac{\pi}{4}, 0, \frac{3 \pi}{4}, 0) $ with $\lambda_1=0.74$ and $\lambda_2=0.89$. We also observe that when $S_1 = 2$ and $S_2 = 2$, the maximum QM value of $S_3$ is $1.88$. Hence, quantum mechanical violation of CFFW inequality (\ref{chshst}) for Bob$^3$ is not possible for any choice of measurement settings when Bob$^1$ and Bob$^2$ get quantum mechanical violation of CFFW inequality (\ref{chshst}). 

It is to be noted here that Bob$^3$ may obtain quantum mechanical violation of the CFFW inequality for steering if the sharpness parameter of Bob$^2$ is too small to get a violation. In fact, it can be easily checked that Alice can be steered by any one pair of the combinations: (Bob$^1$, Bob$^2$), (Bob$^2$, Bob$^3$), (Bob$^1$, Bob$^3$). Therefore, Alice's particle can be steered by two Bobs in terms of the QM violation of the CFFW inequality \cite{achsh}. 
Following a similar procedure, we have further checked that, for the CJWR inequality \cite{cvl}  with two measurements performed by each party, not more than two Bobs can demonstrate steering with a single Alice.

\subsection{C. Three Bobs can steer Alice with three measurements for each party}

We now investigate in the aforementioned scenario, how many Bobs can demonstrate steering with a single Alice if the number of measurements per party is increased. Specifically, we consider the
CJWR inequality with three measurements performed by each party.
In this case too we assume that a singlet state is  shared between Alice and multiple Bobs. 
Suppose, Alice has a choice between three dichotomic measurements: spin component observables in the directions $\{ \hat{x^0}, \hat{x^1}, \hat{x^2} \}$, to perform, and Bob$^n$ also has the choice between spin component observables in the directions $\{ \hat{y^0_n}, \hat{y^1_n},  \hat{y^2_n} \}$. Outcomes of these measurements are labelled by $\{-1, +1 \}$.

We again consider an unbiased input scenario, i.e., all three possible  measurement settings of the previous Bobs are equally probable with probability $\frac{1}{3}$. Hence, the average correlation function between Alice and Bob$^2$, when Alice performs a projective measurement of the spin component along the direction $\hat{x^j}$, and Bob$^2$ performs an unsharp measurement of the spin component along the direction $\hat{y_2^k}$, is given by,
\begin{equation}
\overline{C_2^{jk}} = -\frac{\lambda_2}{3}\Big[\sqrt{1-\lambda_1^2} (\hat{y_2^k} \cdot \hat{x^j}) + \Big(1- \sqrt{1-\lambda_1^2}\Big)  (\hat{y_1^0} \cdot \hat{x^j})  (\hat{y_2^k} \cdot \hat{y_1^0})\Big] \nonumber
\end{equation}
\begin{equation}
-\frac{\lambda_2}{3}\Big[\sqrt{1-\lambda_1^2} (\hat{y_2^k}\cdot \hat{x^j}) + \Big(1- \sqrt{1-\lambda_1^2}\Big)  (\hat{y_1^1} \cdot \hat{x^j})  (\hat{y_2^k} \cdot \hat{y_1^1})\Big] \nonumber
\end{equation}
\begin{equation}
-\frac{\lambda_2}{3}\Big[\sqrt{1-\lambda_1^2} (\hat{y_2^k}\cdot \hat{x^j}) + \Big(1- \sqrt{1-\lambda_1^2}\Big)  (\hat{y_1^2}\cdot \hat{x^j})  (\hat{y_2^k} \cdot \hat{y_1^2})\Big].
\end{equation}
In a similar way, the average correlation function between Alice and Bob$^n$ $\overline{C_n^{jk}}$ can be evaluated.

In this scenario, the left hand side of the CJWR inequality \cite{cvl} for Bob$^n$ to steer Alice's particle is given by,
\begin{equation}
\label{cjwr}
F^3_n = \dfrac{1}{\sqrt{3}} \Big| \sum_{i=1}^{3}\overline{C_n^{ii}} \Big|.
\end{equation}

The three possible choices of measurement settings of Alice and Bob$^n$ are the spin component observables in the directions $\hat{x^i}$ given by Eq.(\ref{alicedir}) and $\hat{y_n^i}$ given by Eq.(\ref{bobndir}) respectively, where $i\in\{0,1,2\}$.
As Alice's system is to be steered, the direction of her spin component measurements should be mutually orthogonal, i. e., $\hat{x^0} \cdot \hat{x^1}$ = $0$; $\hat{x^0} \cdot \hat{x^2}$ = $0$; $\hat{x^1} \cdot \hat{x^2}$ = $0$.

In this case we find that not only Bob$^1$, Bob$^2$, but also Bob$^3$ can steer Alice's particle. Suppose that the measurements of the final Bob (i.e., Bob$^3$) are sharp ($\lambda_3 = 1$), and Bob$^1$, Bob$^2$ measure weakly. In particular, we observe that when each of the QM violations of the CJWR inequality for steering from Bob$^2$ to Alice and that for steering from Bob$^1$ to Alice is $5\%$, i.e., when $F^3_1 = 1.05$ and $F^3_2 = 1.05$, then the maximum QM value of the left hand side of the CJWR inequality for steering from Bob$^3$ to Alice is given by, $F^3_3 = 1.21$. This happens for the choice of measurement settings: ($\theta^x_0$, $\phi^x_0$, $\theta^x_1$, $\phi^x_1$, $\theta^x_2$, $\phi^x_2$, $\theta^{y_1}_0$, $\phi^{y_1}_0$, $\theta^{y_1}_1$, $\phi^{y_1}_1$, $\theta^{y_1}_2$, $\phi^{y_1}_2$, $\theta^{y_2}_0$, $\phi^{y_2}_0$, $\theta^{y_2}_1$, $\phi^{y_2}_1$, $\theta^{y_2}_2$, $\phi^{y_2}_2$, $\theta^{y_3}_0$, $\phi^{y_3}_0$, $\theta^{y_3}_1$, $\phi^{y_3}_1$, $\theta^{y_3}_2$, $\phi^{y_3}_2$) $\equiv$ $(\frac{\pi}{2}$, $0.12$, $\pi$, $3.24$, $\frac{\pi}{2}$, $1.69$, $\frac{\pi}{2}$, $3.26$, $0$, $2.05$, $\frac{\pi}{2}$, $4.83$, $\frac{\pi}{2}$, $0.12$, $\pi$, $5.95$, $\frac{\pi}{2}$, $1.69$, $\frac{\pi}{2}$, $3.26$, $0$, $0.06$, $\frac{\pi}{2}$, $4.83$) with $\lambda_1=0.61$ and $\lambda_2=0.70$. It is observed that when $\lambda_1$ = $0.58$ (i.e., when $F_1^3 = 1$) for $\lambda_2 \in[0.66,0.86]$ both Bob$^2$ and Bob$^3$ can steer Alice. Similarly, when $\lambda_1$ = $0.64$ (i.e., when $F_1^3 = 1.10$) for $\lambda_2 \in[0.68,0.84]$, both Bob$^2$ and Bob$^3$ can steer Alice. Hence, in this case we find that all three Bobs, i. e., Bob$^1$, Bob$^2$ and Bob$^3$  can steer Alice's particle.

We further find that not more than three Bobs can steer Alice's system  through violation of the three settings CJWR inequality.
For four Bobs measuring sequentially, the measurements of the final Bob (i.e., Bob$^4$) are sharp ($\lambda_4 = 1$), and the measurements of Bob$^1$, Bob$^2$ and Bob$^3$ are unsharp. It is observed that when $F^3_1 = 1$, $F^3_2 = 1$  and $F^3_3 = 1$,  the maximum QM value of the left hand side of the CJWR inequality for steering from Bob$^4$ to Alice is given by, $F^3_4 = 0.94$.
It is to be noted here that Bob$^4$ may obtain quantum mechanical violation of the CJWR inequality for steering if the sharpness parameter of any one of the previous Bobs is too small  to get a violation. It can be easily checked that Alice can be steered by any of the  combinations: (Bob$^1$, Bob$^2$, Bob$^3$), (Bob$^1$, Bob$^2$, Bob$^4$), (Bob$^1$, Bob$^3$, Bob$^4$) or (Bob$^2$, Bob$^3$, Bob$^4$). Therefore, Alice's particle can be steered by at most three Bobs in terms of the QM violation of 
the three settings CJWR inequality \cite{cvl}. 

\textbf{Conjecture:} 
In view of the above results, it is legitimate to ask whether the number of Bobs, who can steer Alice, increases upon increasing the number of measurements performed by each party. The CJWR inequality with four, six, ten dichotomic measurements per observer was considered in \cite{nature} with measurement directions along vertices of regular Platonic solids. Here we conjecture that when steering is probed through an $n$-settings CJWR inequality, at most $n$ Bobs can demonstrate steering of Alice's system.

\section{IV. CONCLUSIONS}

In case of monogamy relations for different steering inequalities, $n$ observers share $n$ number of particles, one particle per each observer, and all the observers are spatially separated, and hence, the no-signalling condition is obeyed by each pair of observers sharing the state as a consequence of relativistic causality. In the present study we consider a scenario where Alice accesses half of an entangled pair and multiple Bobs access and perform measurements sequentially on the other half. In this case, Alice and multiple Bobs are spatially separated and the no-signalling condition is satisfied between Alice and any Bob. On the other hand, multiple Bobs perform 
measurements on the same particle sequentially. By the act of
his measurement, each Bob signals to the subsequent Bob who accesses the given particle. This is why no-signalling is not physically relevant between different Bobs. We consider that each Bob performs measurements, independent of the measurement settings and outcomes of the previous Bobs on the particle in his possession and we also consider an unbiased input scenario. In this scenario, we find that at most two Bobs can demonstrate violation of the CFFW inequality \cite{achsh}, or violation of the two settings CJWR inequality \cite{cvl} with a single Alice  going beyond the monogamy restriction. We also show that at most three Bobs can steer Alice's system, if we consider the three settings CJWR inequality.
 We further conjecture that when steering is probed through an $n$-settings CJWR inequality, at
 most $n$ Bobs can demonstrate steering of Alice's system.
There are some open questions related to this study. First, it remains open to prove or disprove the conjecture we have made here. It was shown in previous works \cite{sygp, mal} that the QM violation of the CHSH inequality can be shared by at most two Bobs with a single Alice by using two dichotomic measurements per party. Hence, it will be interesting to find whether more than two Bobs can share nonlocality with a single Alice considering violation of other Bell-type inequalities in the context of more number of measurement settings per party. 

\section{ ACKNOWLEDGEMENTS}
 S. S. acknowledges the financial support from INSPIRE programme, Department of Science and Technology, Government of India. D. D. acknowledges the financial support from University Grants Commission (UGC), Government of India.

\end{document}